\documentclass[structabstract]{aa}  
\usepackage{graphicx}
\usepackage{txfonts}
\usepackage{natbib}
\bibpunct{(}{)}{;}{a}{}{,}
\usepackage{url}

\newcommand{\unit}[1]{\ensuremath{\,\mathrm{#1}}}

\newcommand{\corotthanks}{\thanks{The CoRoT space mission, launched on December 27th 2006, has been developed and is operated by CNES, with the contribution of Austria, Belgium, Brazil, ESA (RSSD and Science Programme), Germany and  Spain.}}

\newcommand{\teff}{\ensuremath{T_{\mathrm{eff}}}}
\newcommand{\vmicro}{\ensuremath{v_{\mathrm{micro}}}}
\newcommand{\logg}{\ensuremath{\log g}}
\newcommand{\feh}{\ensuremath{\mathrm{[Fe/H]}}}
\newcommand{\meh}{\ensuremath{\mathrm{[M/H]}}}
\newcommand{\vsini}{\ensuremath{v \sin i}}
\newcommand{\kms}{\unit{km\,s^{-1}}}
\newcommand{\maa}{m\AA}
\newcommand{\hd}{HD~52265}
\newcommand{\dc}{\ensuremath{r_{\mathrm{dc}}}}
\newcommand{\muHz}{\unit{\mu Hz}}
\newcommand{\Dnu}{\ensuremath{\overline{\Delta\nu}}}
\begin{document}

\title{Accurate p-mode measurements of the G0V metal-rich CoRoT\corotthanks\ target \hd}

\author{%
J.~Ballot\inst{\ref{inst:irap1},\ref{inst:irap2}} \and
L.~Gizon\inst{\ref{inst:mps}} \and
R.~Samadi\inst{\ref{inst:lesia}} \and
G.~Vauclair\inst{\ref{inst:irap1},\ref{inst:irap2}} \and
O.~Benomar\inst{\ref{inst:ias},\ref{inst:sifa}} \and
H.~Bruntt\inst{\ref{inst:lesia}} \and
B.~Mosser\inst{\ref{inst:lesia}} \and
T.~Stahn\inst{\ref{inst:mps}} \and
G.~A.~Verner\inst{\ref{inst:auqm},\ref{inst:spub}} \and 
T.~L.~Campante\inst{\ref{inst:dasc},\ref{inst:caup}} \and
R.~A.~Garc\'\i a\inst{\ref{inst:aim}} \and
S.~Mathur\inst{\ref{inst:hao}} \and 
D.~Salabert\inst{\ref{inst:ull},\ref{inst:iac},\ref{inst:nice}} \and
P.~Gaulme\inst{\ref{inst:ias}} \and
C.~R\'egulo\inst{\ref{inst:ull},\ref{inst:iac}} \and 
I.~W.~Roxburgh\inst{\ref{inst:auqm}} \and
T.~Appourchaux\inst{\ref{inst:ias}} \and
F.~Baudin\inst{\ref{inst:ias}} \and
C.~Catala\inst{\ref{inst:lesia}} \and
W.~J.~Chaplin\inst{\ref{inst:spub}} \and
S.~Deheuvels\inst{\ref{inst:lesia}} \and
E.~Michel\inst{\ref{inst:lesia}} \and
M.~Bazot\inst{\ref{inst:caup}} \and
O.~Creevey\inst{\ref{inst:ull},\ref{inst:iac}} \and
N.~Dolez\inst{\ref{inst:irap1},\ref{inst:irap2}} \and
Y.~Elsworth\inst{\ref{inst:spub}} \and
K.~H.~Sato\inst{\ref{inst:aim}} \and
S.~Vauclair\inst{\ref{inst:irap1},\ref{inst:irap2}} \and
M.~Auvergne\inst{\ref{inst:lesia}} \and
A.~Baglin\inst{\ref{inst:lesia}}
}

  \institute{%
Institut de Recherche en Astrophysique et Plan\'etologie, CNRS, 14 avenue Edouard Belin, 31400 Toulouse, France\label{inst:irap1}\\
\email{jballot@ast.obs-mip.fr}
\and
Universit\'e de Toulouse, UPS-OMP, IRAP, Toulouse, France\label{inst:irap2}
\and
Max-Planck-Institut f\"ur Sonnensystemforschung, 37191 Katlenburg-Lindau, Germany\label{inst:mps}
\and
LESIA, UMR8109, Universit\'e Pierre et Marie Curie, Universit\'e Denis Diderot, Observatoire de Paris, 92195 Meudon, France\label{inst:lesia}
\and 
Institut d'Astrophysique Spatiale, CNRS, Universit\'e Paris XI, 91405 Orsay, France\label{inst:ias}
\and
Sydney Institute for Astronomy, School of Physics, University of 
Sydney, NSW 2006, Australia\label{inst:sifa} 
\and
Astronomy Unit, Queen Mary, University of London Mile End Road, London E1 4NS, UK\label{inst:auqm}
\and
School of Physics and Astronomy, University of Birmingham, Edgbaston, Birmingham B15 2TT, UK\label{inst:spub}
\and 
Danish AsteroSeismology Centre, Department of Physics and Astronomy, University of Aarhus, 8000 Aarhus C, Denmark\label{inst:dasc}
\and
Centro de Astrof\'isica, DFA-Faculdade de Ci\^encias, Universidade do  
Porto, Rua das Estrelas, 4150-762 Porto, Portugal\label{inst:caup}
\and
Laboratoire AIM, CEA/DSM, CNRS, Universit\'e Paris Diderot, IRFU/SAp, Centre de Saclay, 91191 Gif-sur-Yvette Cedex, France\label{inst:aim}
\and
High Altitude Observatory, NCAR, P.O. Box 3000, Boulder, CO 80307, USA\label{inst:hao}
\and
Universidad de La Laguna, Dpto de Astrof\'\i sica, 38206, La Laguna, Tenerife, Spain\label{inst:ull}
\and
Instituto de Astrof\'\i sica de Canarias, 38205, La Laguna, Tenerife, Spain\label{inst:iac}
\and
Universit\'e de Nice Sophia-Antipolis, CNRS, Observatoire de la C\^ote d'Azur, BP 4229, 06304 Nice Cedex 4, France\label{inst:nice}
}

\date{Received 20 January 2011 / Accepted 22 March 2011}

  \abstract%
{The star \object{\hd} is a G0V metal-rich exoplanet-host star observed in the seismology field of the CoRoT space telescope from November 2008 to March 2009. The satellite collected 117 days of high-precision photometric data on this star, showing that it presents solar-like oscillations. \hd\ was also observed in spectroscopy with the Narval spectrograph at the same epoch.}%
{We characterise \hd\ using both spectroscopic and seismic data. }%
{The fundamental stellar parameters of \hd\ were derived with the semi-automatic software VWA, and the projected rotational velocity was estimated by fitting synthetic profiles to isolated lines in the observed spectrum. The parameters of the observed p modes were determined with a maximum-likelihood estimation. We performed a global fit of the oscillation spectrum, over about ten radial orders, for degrees $l=0$ to 2. We also derived the properties of the granulation, and analysed a signature of the rotation induced by the photospheric magnetic activity.}%
{Precise determinations of fundamental parameters have been obtained: $\teff=6100 \pm 60\unit{K}$, $\log g=4.35  \pm 0.09$, $\meh=0.19  \pm 0.05$, as well as $\vsini=3.6^{+0.3}_{-1.0}\kms$. We have measured a mean rotation period $P_\mathrm{rot}=12.3\pm0.15$ days, and find a signature of differential rotation. The frequencies of 31 modes are reported in the range 1500--2550\muHz. The large separation exhibits a clear modulation around the mean value $\Dnu=98.3\pm0.1\muHz$. Mode widths vary with frequency along an S-shape with a clear local maximum around 1800\muHz. We deduce lifetimes ranging between 0.5 and 3 days for these modes. Finally, we find a maximal bolometric amplitude of about $3.96\pm0.24$~ppm for radial modes.}%
{}%


\titlerunning{Accurate p-mode measurements of \hd}
\maketitle

\section{Introduction}

Late-type stars oscillate when they have sufficiently deep convective envelopes. These oscillations are p modes, i.e., acoustic waves trapped in the stellar interior. Despite being normally stable, these modes are stochastically excited by near-surface turbulent motions \citep[e.g.][]{Goldreich77}. Their amplitudes remain small and produce tiny luminosity fluctuations. Observing them is now possible thanks to high-performance dedicated instrumentation such as space-borne photometry.
The current space-based missions CoRoT \citep[Convection, Rotation, and planetary Transits,][]{Baglin06} and Kepler \citep{Koch10} provide long uninterrupted time series of high-precision photometric data over months or years, allowing for the detection and the characterisation of p modes in late-type stars \citep[e.g.][]{Michel08,Gilliland10}. These high-quality data give an exciting opportunity to conduct seismological analyses of solar-like stars.
CoRoT has been successful in providing observations that allow detailed seismic analysis of late-type main-sequence 
\citep[e.g.][]{Appourchaux08,Barban09,Garcia09,Mosser09,Benomar09,Mathur10,Deheuvels10} and giant \citep[e.g.][]{deRidder09,Hekker09,Mosser10} stars showing solar-like oscillations.
Such accurate asteroseismic observations provide unprecedented constraints on the stellar structure of these classes of stars \citep[e.g.][]{DeheuvelsM10,Miglio10}.
The Kepler mission is also currently providing interesting results on these two categories of stars \citep[e.g.][]{Chaplin10,Huber10,Kallinger10,Mathur11}.

In this paper, we study \hd, a G0, rather metal-rich, main-sequence star hosting a planet that was independently discovered in 2000 by \citet{Butler00} and \citet{Naef01}. \hd\ was observed by CoRoT during almost four months from 13 November 2008 to 3 March 2009. CoRoT observations were  complemented by ground-based observations achieved with the Narval spectrograph at the Pic du Midi observatory in December 2008 and January 2009, i.e. during the CoRoT observations.

Up to now, only a very few exoplanet-host stars (EHS) have been studied seismically. Ground-based observations of $\mu$~Ara, hosting two known planets, allowed \citet{Bouchy05} to extract about 30 modes based on an eight-night observation and constrain its fundamental parameters \citep{Bazot05,Soriano10}. Recently, similar constraints have been placed on another EHS, $\iota$~Hor \citep{Vauclair08}. CoRoT has also observed HD~46375, an unevolved K0 star hosting a Saturn-like planet, for which a mean large separation has been measured \citep{Gaulme10}. Kepler also observed the EHS HAT-P-7, for which frequencies of 33 p modes were measured within 1.4\muHz\ accuracy, as well as the EHS HAT-P-11 and TrES-2, for which large separations have been estimated \citep{JCD10}.
The opportunities to get seismic constraints on EHS are then rare. The high-quality observations of \hd\ allow us to determine its seismic properties with an accuracy never obtained on EHS.

In Sect.~\ref{sec:glob}, we first report the fundamental parameters of the star obtained with our spectroscopic study, including a detailed composition analysis. After describing the CoRoT photometric data in Sect.~\ref{sec:corot}, we measure the rotation period and discuss the activity of the star (Sect.~\ref{sec:act}). We also measure granulation properties (Sect.~\ref{sec:bg}) and extract the main p-mode characteristics -- frequencies, amplitudes, and lifetimes -- in Sect.~\ref{sec:pmode}, before concluding (Sect.~\ref{sec:conclusion}).

\section{Fundamental stellar parameters}\label{sec:glob}
\subsection{Distance, luminosity, and chromospheric activity}\label{ssec:distlumact}
The star \hd\ (or HIP~33719, or HR~2622) was initially classified as G0III-IV in the Bright Star Catalogue \citep{Hoffleit82_BSC} before finally being identified as a G0V dwarf thanks to Hipparcos parallax measurements \citep[see][]{Butler00}. It has a magnitude $V=6.301$ and a parallax of $\pi = 34.54 \pm 0.40\unit{mas}$
in the revised Hipparcos catalogue by \citet{vanLeeuwen07} and this corresponds to a distance $d=28.95 \pm 0.34\unit{pc}$.

The catalogue of \citet{vanLeeuwen07} also gives the magnitude $HP=6.4132 \pm 0.0006$. Using the bolometric correction $BC_{HP}= -0.17\pm 0.01$ after \citet{Cayrel97}, one finds an absolute bolometric magnitude $M_{\mathrm{bol}} = 3.93 \pm 0.06$; taking $M_{\mathrm{bol},\sun}=4.75$, this gives  luminosity of $L/L_{\sun}=2.09\pm 0.24$.
For comparison, we also derived the luminosity by using the $V$ magnitude and the bolometric correction $BC_V=-0.03\pm 0.01$ \citep{Flower96} and obtained a fully consistent value for $L$.

The magnetic activity of a star can be revealed by chromospheric emission lines, especially the very commonly used \ion{Ca}{ii} H and K lines, from which is derived the standard activity index $R'_{\mathrm{HK}}$. For \hd, \citet{Wright04} have measured $\log R'_{\mathrm{HK}}=-5.02$, which is consistent with the value of $\log R'_{\mathrm{HK}}=-4.99$ obtained by \citet{Butler00}. These values, comparable to the solar one, indicate \hd\ can be classified as a magnetically quiet star.
These parameters are summarised in Table~\ref{tab:global}.

\begin{table}
\centering
\caption{Summary of general parameters of \hd. See text for details.\label{tab:global}}
\begin{tabular}{lrl}
\hline\hline
 & & Ref. \\
\hline
$\pi$       & $34.54 \pm 0.40\unit{mas}$ & 1\\
$L/L_{\sun}$& $2.09\pm 0.24$             & 1,2\\
$\log R'_{\mathrm{HK}}$ & $-5.02$      & 3\\
\hline
\teff     &  $6100 \pm 60\unit{K}$ & 4\\
$\log g$  &  $4.35  \pm 0.09$ & 4 \\
\meh      &  $0.19  \pm 0.05$ & 4 \\
\vsini    &  $3.6^{+0.3}_{-1.0}\kms$ & 4\\
\hline
\end{tabular}
\tablebib{(1) \citet{vanLeeuwen07}; (2) \citet{Cayrel97}; (3) \citet{Wright04}; (4) this work.}
\end{table}

\subsection{Narval spectroscopic observations}\label{ssec:spectro}

Complementary ground-based observations were obtained with the Narval\footnote{\url{http://www.ast.obs-mip.fr/projets/narval/v1/}} 
spectropolarimeter installed on the Bernard Lyot Telescope at the Pic du Midi Observatory (France).
Spectra with a resolution of 65\,000 were registered simultaneously in classical spectroscopy (Stokes $I$) and circularly polarised light (Stokes $V$) over the spectral domain 370-1000~nm in a single exposure.
Nine high signal-to-noise ratio spectra were obtained in December 2008 and January 2009, i.e., during the CoRoT observation for this star: 1 spectrum recorded on 2008/12/20, 1 on 2008/12/21, 2 on 2009/1/10, and 5 on 2009/1/11.
No Zeeman signature has been recovered in the polarised spectra. This gives us an upper limit of $\sim$1~G for the average magnetic field by combining the five spectra registered on 2009/1/11 and $\sim$2~G for the other days.

We analysed the observed spectrum of \hd\ using the semi-automatic
software package VWA \citep{Bruntt04, Bruntt09}. 
We adopted atmospheric models interpolated in the MARCS grid 
\citep{Gustafsson08} and atomic parameters from VALD \citep{kupka99}.
The abundances for more than 500 lines were 
calculated iteratively by fitting synthetic profiles 
to the observed spectrum using SYNTH \citep{Valenti96}.
For each line, the abundances were calculated differentially with respect
to the same line in a solar spectrum from \citet{Hinkle00}.
The atmospheric parameters (\teff, \logg, \feh, and \vmicro) were
determined by requiring that Fe lines give the same abundance 
independent of equivalent width (EW), excitation potential or
ionisation stage. Only weak lines were used (${\rm EW}<90$\,\maa) for
this part of the analysis, but stronger lines (${\rm EW}<140$\,\maa) were
used for the calculation of the final mean abundances. 
The uncertainties on the atmospheric parameters and the abundances 
were determined by perturbing the best-fitting atmospheric 
parameters as described by \citet{Bruntt08}. 

\begin{figure*}[!ht]
\includegraphics[height=.8\textwidth,angle=90]{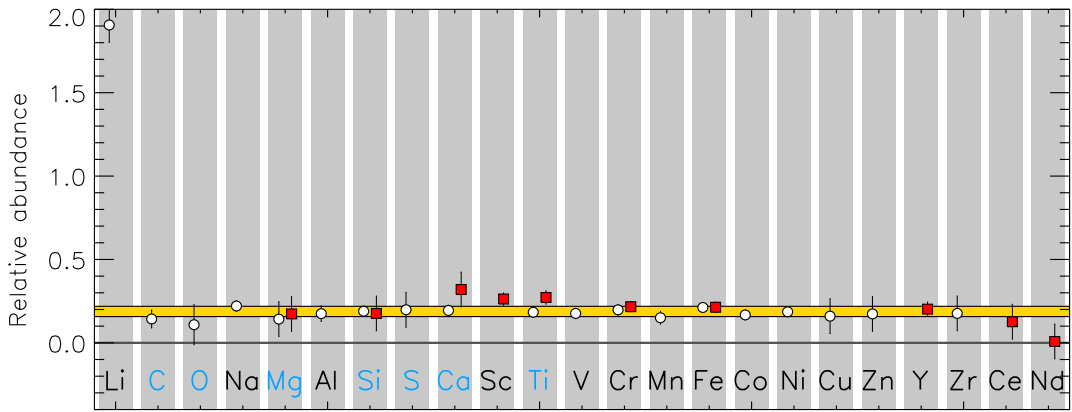}
 \caption{Abundance pattern of \hd\ for 23 elements. Circles and box symbols are used for the mean abundance from neutral and singly ionised lines, respectively. The yellow horizontal bar marks the mean metallicity with 1-$\sigma$ uncertainty range. The chemical symbols in blue correspond to $\alpha$ elements.}
\label{fig:abund} 
\end{figure*}

The results are 
$\teff =  6100 \pm 31 \pm 50$~K,
$\log g = 4.35 \pm 0.04 \pm 0.08$,
$\vmicro = 1.18 \pm 0.04 \pm 0.05\kms$.
We give two uncertainties: the first one is the intrinsic error
and the second is our estimated systematic error. The systematic errors
were estimated from a large sample of F5-K1 type stars \citep{Bruntt10}.
These two errors are independent and should be added quadratically.

We have also fitted profiles to 68 isolated lines to find the best combination of \vsini\ and macroturbulence. Lines with high $\chi^2$ of the best fit have been discarded, and 49 of the best lines remain. We found $\vsini= 3.6^{+0.3}_{-1.0}\kms$.

We calculated the mean metallicity from the six metals
with at least 10 lines (Si, Ti, V, Cr, Fe, and Ni): 
$\meh= +0.19 \pm 0.05$, where the uncertainty includes the errors on
the atmospheric parameters.
In Table~\ref{tab:abund} we list the abundances of the 23 elements
that are also shown in Fig.~\ref{fig:abund}. The estimated internal error is 0.04 dex for all elements.

\begin{table}[!htb]
\centering
\caption{Abundances in \hd\ relative to the Sun ($\log \epsilon/\epsilon_{\sun}$) and number of lines ($N$) used in spectral analysis.}
\label{tab:abund}
\begin{tabular}{lcr|lcr}
\hline
\hline
Elem. & $\log \epsilon/\epsilon_{\sun}$ & $N$ &
Elem. & $\log \epsilon/\epsilon_{\sun}$ & $N$  \\
      &  [dex]                          &     &
      &  [dex]                          &       \\
\hline
  \ion{Li}{i} &  1.91  &   1  &  \ion{V}{i}  &  0.18  &  12  \\  
  \ion{C}{i}  &  0.14  &   6  &  \ion{Cr}{i} &  0.20  &  21  \\  
  \ion{O}{i}  &  0.11  &   3  &  \ion{Cr}{ii}&  0.22  &   4  \\  
  \ion{Na}{i} &  0.22  &   4  &  \ion{Mn}{i} &  0.15  &   7  \\  
  \ion{Mg}{i} &  0.14  &   2  &  \ion{Fe}{i} &  0.19  & 268  \\  
  \ion{Mg}{ii}&  0.17  &   2  &  \ion{Fe}{ii}&  0.21  &  23  \\  
  \ion{Al}{i} &  0.17  &   3  &  \ion{Co}{i} &  0.17  &   9  \\  
  \ion{Si}{i} &  0.19  &  34  &  \ion{Ni}{i} &  0.19  &  72  \\  
  \ion{Si}{ii}&  0.18  &   2  &  \ion{Cu}{i} &  0.16  &   2  \\  
  \ion{S}{i}  &  0.20  &   2  &  \ion{Zn}{i} &  0.17  &   2  \\ 
  \ion{Ca}{i} &  0.19  &  12  &  \ion{Y}{ii} &  0.20  &   4  \\ 
  \ion{Ca}{ii}&  0.32  &   1  &  \ion{Zr}{i} &  0.18  &   1  \\  
  \ion{Sc}{ii}&  0.26  &   6  &  \ion{Ce}{ii}&  0.13  &   1  \\  
  \ion{Ti}{i} &  0.18  &  27  &  \ion{Nd}{ii}&  0.01  &   1  \\  
  \ion{Ti}{ii}&  0.27  &  10  &              &        &      \\
\hline  
\end{tabular}
\end{table}

\hd\ is relatively metal rich ($\approx50$\% more heavy elements than the Sun),
but there is no evidence of any strong enhancement of $\alpha$ elements.
As seen in Fig.~\ref{fig:abund}, almost all elements are enhanced by a similar amount. We can then use a global scaling of the metallicity of $+0.19$. 
Nevertheless, it is seen that the star has a high lithium abundance ($1.91\unit{dex}$). For the region around the \ion{Li}{i} line at 6707.8\,\AA\ we used the line list from \cite{Ghezzi09} but  did not include the CN molecular lines. Considering the abundance
$\log\epsilon(\ion{Li})_{\sun}=1.05 \pm 0.1$ for the Sun 
\citep{Asplund09}, we got $\log\epsilon(\ion{Li})=2.96 \pm 0.15$ for \hd.
This measure is consistent with results of \citet{Israelian04}. According to their sample, such a high abundance of lithium is not unusual for a star with $\teff=6100\unit{K}$.

The fundamental parameters are summarised in Table~\ref{tab:global}. These were compared to previous work, especially to \citet{Valenti05}. We recover consistent results for \teff, \meh, and \logg, within the error bars. The major change concerns \vsini, which is significantly reduced from $4.7 \pm 0.5\kms$ to $3.6^{+0.3}_{-1.0}\kms$, probably owing to the inclusion of macroturbulence in the present work.

\section{CoRoT photometric observations}\label{sec:corot}

\subsection{CoRoT data} \label{ssec:data}
\hd\ was observed by CoRoT in the seismology field \citep[see][]{Auvergne09} during 117 days starting on 2008 November 13 to 2009 March 3, during the second long run in the galactic anti-centre direction (LRa02). We used time series that are regularly spaced in the heliocentric frame with a 32~s sampling rate, i.e. the so-called Helreg level 2 (N2) data \citep{Samadi07astroph}. The overall duty cycle of the time series is $\dc=94.6\%$. However, the extra noise on data taken across the south-Atlantic anomaly (SAA) creates strong harmonics of the satellite orbital period and of the day \citep[for more details, see][]{Auvergne09}. When all of the data registered  across the SAA are removed, the duty cycle becomes $\dc=90.2\%$.

\begin{figure}[!htb]
  \centering
  \includegraphics[width=\linewidth]{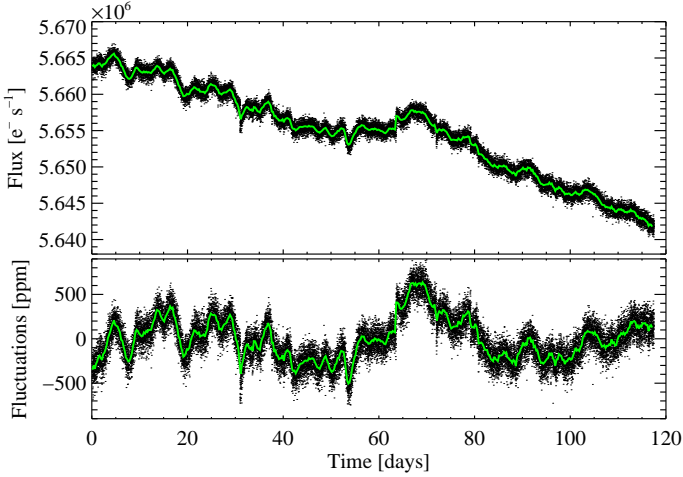} 
  \caption{\emph{(top)} Light curve of \hd\ observed by CoRoT, starting on 2008/11/13. Points taken across the SAA have been removed. The green curve is resampled within 8-hour bins. \emph{(bottom)} The fluctuations around the average obtained after detrending with a 3rd-order polynomial.}
  \label{fig:lc}
\end{figure}

The light curve of \hd\ from N2 data is plotted in Fig.~\ref{fig:lc}-\emph{top}. The light curve shows a slowly decreasing trend due to instrumental effects, and especially due to the ageing of CCDs. This trend is removed with a third-order polynomial fit. Residual fluctuations are plotted in the bottom of Fig.~\ref{fig:lc}.

The power spectral density (PSD) is computed with a standard fast Fourier transform (FFT) algorithm, and we normalised it as the so-called one-sided power spectral density \citep{NumRec}. We interpolate the gaps produced by SAA with parabola fitting the points around the gap, in a similar way to the usual interpolation performed in N2 data for missing points.

To verify that the interpolation method has no influence on the results presented in this paper, we performed similar analyses on spectra where:
\begin{enumerate}
 \item the PSD is computed as the FFT of the N2 data, and main harmonics and aliases induced by the SAA noise are removed in the spectrum;
\item all of the gaps and missing points are interpolated with the in-painting method described in
\citet{Mathur10} and \citet{Sato10}.
\end{enumerate}
We have also verified that the order of the polynomial used to detrend the data has no impact on the results, by using first to ninth order polynomials.

\subsection{Influence of the observation window}\label{ssec:window}
At the precision reached by the current analysis, windowing effects must be taken into account when normalising the PSD.
The first consequence of windowing is a lack of power in the PSD by a factor of \dc. This factor should be taken into account a posteriori to avoid a general underestimation of the power by around 10\%.
Of course, the interpolation we performed injected power into the spectrum, but only at very low frequency. Thus, the PSD is properly calibrated by dividing it by \dc.

\begin{figure}[!htb]
  \centering
  \includegraphics[width=\linewidth]{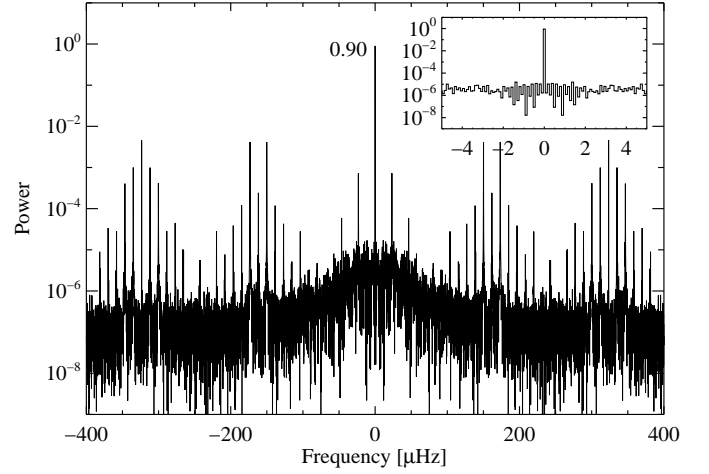}
  \caption{Power spectrum of the observation window with a zoom around the zero frequency in the insert.}
  \label{fig:window}
\end{figure}

Moreover, owing to the repetitive pattern of the gaps, the power is spread in side lobes. To estimate the resulting leakage, we computed the power spectrum of the observation window  (Fig.~\ref{fig:window}).
On a logarithmic scale we clearly see a forest of peaks corresponding to the orbital frequency (161.7\muHz), its harmonics, as well as daily aliases. These peaks have lower amplitudes than the central peak (smaller than 1\%), and there is no peak close to the central one. As a result, all of the p modes, which are narrow structures, should be modelled well by Lorentzian profiles (see Sect.~\ref{sec:pmode}). The mode aliases then have very small amplitudes and do not perturb the other modes significantly, but a non-negligible part of the mode power is spread in the background, and the fitted mode heights will be underestimated by a factor of $\dc$.

However, we neglect the leakage effects on very broad structures, such as the background ones (see Sect.~\ref{sec:bg}), because their widths are at least as broad as the range where the significant peaks of the window spectrum are found.

\section{Activity, spots, and rotation period}\label{sec:act}
The light curve (Fig.~\ref{fig:lc}) exhibits clear variations on characteristic time scales of a few days and with amplitudes around 100 ppm, which we attribute to stellar activity. As shown in Sect.~\ref{ssec:distlumact}, this star is magnetically quiet, much like the Sun. The amplitude of the modulation is then compatible with those produced by spots in the photosphere. The modulation appears quasi-periodic with clear repetitive patterns.

\begin{figure}[!htb]
  \centering
  \includegraphics[width=\linewidth]{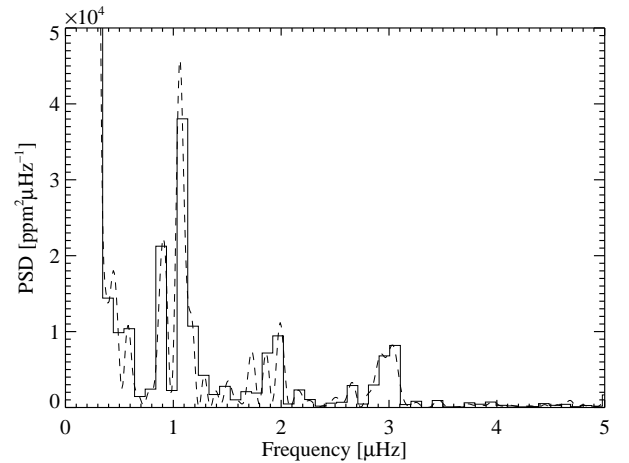}
  \caption{PSD of \hd\ at low frequency. Dashed line shows an oversampling of the PSD by a factor of 10.}
  \label{fig:spectrum}
\end{figure}

This modulation produces a significant peak in the power spectrum at 1.05\muHz\ and a second very close one, at 0.91\muHz\ (Fig.~\ref{fig:spectrum}). The signature is then not a single peak, but a slightly broadened structure. We interpret the broadening as the signature of differential rotation, since spots may appear at different latitudes that rotate with different periods.
Peaks around 2 and 3\muHz\ are simply the harmonics of the rotation frequency.

\begin{figure}[!htb]
  \centering
  \includegraphics[width=\linewidth]{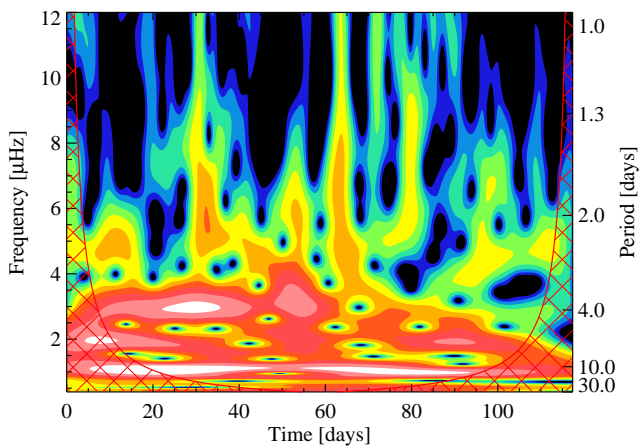}
  \caption{Wavelet power spectrum for \hd\ as a function of the frequency of the Morlet wavelet (y-axis) and time (x-axis). The colour scale is logarithmic, white corresponds to the highest values, and red hatching indicates the cone of influence delimiting the reliable periodicity.}
  \label{fig:wavelet}
\end{figure}

A wavelet analysis \citep[e.g.][]{Torrence98} of the curve was performed to verify the persistence of this characteristic frequency during the complete length of observation. 
Interestingly, this method allowed us to monitor the evolution with time of the power spectral distribution, as well as to resolve the uncertainty between the rotation period and the first harmonic that could be observed in the power spectrum \citep[see][]{Mathur10pipe}.
We used the Morlet wavelet (a moving Gaussian envelope convolved with a sinusoid function with a varying frequency) to produce the wavelet power spectrum
shown Fig.~\ref{fig:wavelet}. It represents the correlation between the wavelet with a given frequency along time. The frequency around 1\muHz\ previously identified as the rotation rate is very persistent during the whole time series, while the peaks at 2 and 3\muHz\ are not. Thus, the 1\muHz\ signature does not come from strong localised artefacts, but is the consequence of a continuous modulation. This confirms that the rotation period is around 12 days.

\begin{figure}[!htb]
  \centering
  \includegraphics[width=\linewidth]{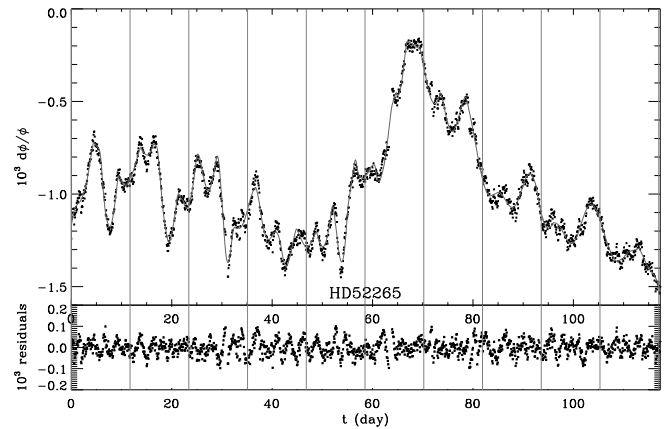}
  \caption{Spot modelling of \hd. Relative luminosity fluctuations ($\mathrm{d}\phi/\phi$) of \hd\ are plotted as a function of time. The dots in the upper panel represent the data binned every CoRoT orbit (6184 s) and the solid grey curve is the best-fit model. Dots in the bottom panel represent residuals. Vertical grey lines indicate the equatorial rotation period.}
\label{fig:spot}
\end{figure}

Finally, we performed spot-modelling of the light curve following 
the method described by \citet{Mosser09spot}. The best-fit model is presented in Fig.~\ref{fig:spot}. The fit is obtained with an inclination of $i=30\pm10\degr$. When we assumed a solid rotation, we found a rotation period $P_\mathrm{rot}=12.3 \pm 0.15$ days. However, considering differential rotation, with a law
\begin{equation}
 P_\mathrm{rot}(\vartheta)=P_\mathrm{eq}/(1-K\sin^2\vartheta)
\end{equation}
for the rotation as a function of the latitude $\vartheta$, provides a better fit. We derive an equatorial rotation period $P_\mathrm{eq}=11.7_{-0.2}^{+0.6}$~days and a differential rotation rate $K = 0.25_{-0.20}^{+0.05}$. The spot life time is short, around $\tau_\mathrm{spot}=8.0\pm 1.5$ days, i.e. about 2/3 of the rotation period. As for similar stars, the modelling is not able to reproduce the sharpest features of the light curve. Fitting them would require introducing too many spots, which lowers the precision of the best-fit parameters. Moreover, some of these features, such as the jumps around $t=32$, 54, or 64 days, certainly do not have a stellar origin, but are probably introduced by instrumental glitches or cosmic events.

\section{Fitting the stellar background}\label{sec:bg}

\begin{figure}[!htb]
 \includegraphics[width=\linewidth]{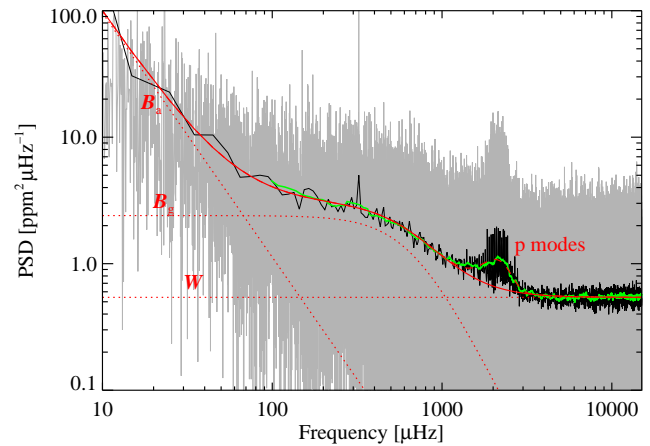}
\caption{Power spectral density of \hd\ at full resolution (grey curve) and rebinned with a factor of 100 (black curve). The green line shows the spectrum smoothed by a box car with a width equal to the mean large separation \Dnu. The fitted background is the solid red line, and its three components ($W$, $B_\mathrm{g}$, $B_\mathrm{a}$, see text) are plotted as red dotted lines. P-mode power excess, fitted as a Gaussian profile, is also represented with a red dashed line.\label{fig:bg}}
\end{figure}

\begin{figure*}[!htb]
 \includegraphics[width=\linewidth]{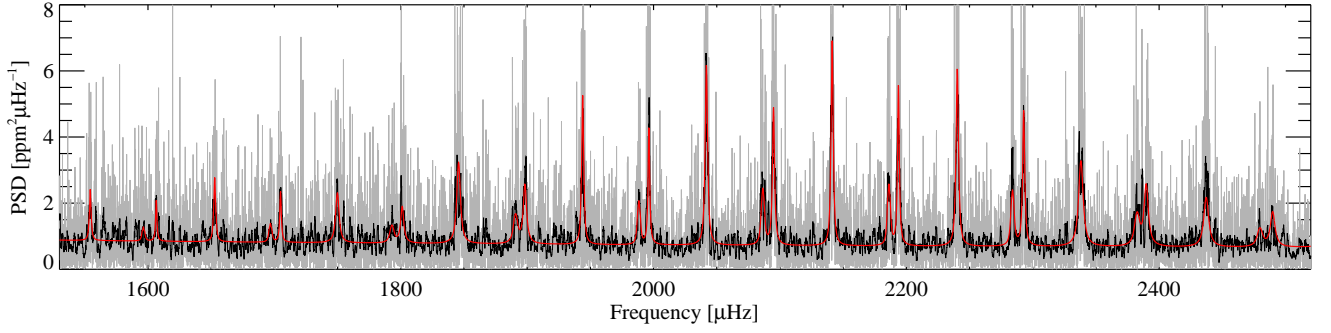}
\caption{Power spectral density of \hd\ in the p-mode frequency range at full resolution (grey curve) and smoothed by a 11-bin wide boxcar (black curve). The red line corresponds to the fitted spectrum.}
\label{fig:fit}
\end{figure*}

The PSD of \hd\ is shown in Fig.~\ref{fig:bg}. The p modes are clearly visible around 2000\muHz. The dominant peak at $323\muHz$ is the persisting first harmonic of the satellite orbital period. Before analysing the p-mode pattern, we determine the properties of the background, which gives useful information about the stellar granulation that constrains, for example, 3-D models of superficial convection \citep[e.g.][]{Ludwig09}.

The background $B(\nu)$ of the spectrum is modelled as the sum of three components:
\begin{enumerate}
 \item a white noise $W$ modelling the photon shot noise, dominating the spectrum at high frequency;
\item a profile of \citet{Harvey85} to model the granulation spectrum:
\begin{equation}
 B_\mathrm{g}(\nu) = \frac{4\tau_\mathrm{g}\sigma_\mathrm{g}^2}{1+(2\pi\tau_\mathrm{g}\nu)^{\alpha_\mathrm{g}}} \label{eq:gran}
\end{equation}
where $\tau_\mathrm{g}$ and $\sigma_\mathrm{g}$ are characteristic time scale and amplitude of the granulation and $\alpha_\mathrm{g}$ an exponent characterising the temporal coherence of the phenomenon;
\item a third component taking the slow drifts or modulations due to the stellar activity, the instrument, etc., into account which we modelled either by an extra Harvey profile $B_\mathrm{a}(\nu)=4\tau_\mathrm{a}\sigma_\mathrm{a}^2/[1+(2\pi\tau_\mathrm{a}\nu)^{\alpha_\mathrm{a}}]$ or a power law $B_\mathrm{a}(\nu)=P_\mathrm{a}\nu^{-e_\mathrm{a}}$.
\end{enumerate}

We fit the PSD between 1\muHz\ and the Nyquist frequency, excluding the region where p modes are visible ([1200-3200]\muHz). Fittings were performed with a maximum-likelihood estimation (MLE), or alternatively with Markov chains Monte Carlo (MCMC), by assuming that the noise follows a $\chi^2_2$ distribution. Results for $B_\mathrm{g}$ and $W$ do not depend significantly on the choice made for $B_\mathrm{a}$. Table~\ref{tab:bg} lists the fitted parameters with statistical formal errors, obtained by inverting the Hessian matrix. These errors do not include any systematics and assume the model is correct.

\begin{table}[!htb]
 \centering
\caption{White noise ($W$) and parameters of the granulation background (Eq.~\ref{eq:gran}) fitted on \hd\ power spectrum. \label{tab:bg}}
\begin{tabular}{cccc}
\hline\hline

$W$&
$\sigma_\mathrm{g}$&
$\tau_\mathrm{g}$&
$\alpha_\mathrm{g}$ \\
$\mathrm{[ppm^2\,\mu Hz^{-1}}]$ &
[ppm] &
[s] &
\\
\hline

$0.542\pm 0.002$ &
$49.8 \pm 0.4$ &
$242 \pm 2$ &
$2.65 \pm 0.15$ \\
\hline
\end{tabular}
\end{table}

Complementary fits including the p-mode region were performed by modelling their contribution as a Gaussian function. The results for $B_\mathrm{g}$ and $W$ have not been modified. The p-mode Gaussian profile $B_\mathrm{p}$ is characterised by a height $H_\mathrm{p}^\mathrm{(max)}=0.426\pm 0.016\unit{ppm^2\,\mu Hz^{-1}}$, a width $\Delta_\mathrm{p}=390\pm20\muHz$, and a central frequency $\nu_\mathrm{p}^\mathrm{(max)}=2090\pm20\muHz$. We then redid fits over a frequency range starting at 5, 10, or 100\muHz, instead of 1\muHz, without any impact on the results. We  also verified that removing the two bins still contaminated by the strong orbit harmonic does not influence the results.

\section{P-mode analysis}\label{sec:pmode}

\begin{figure}[!htb]
 \includegraphics[width=\linewidth]{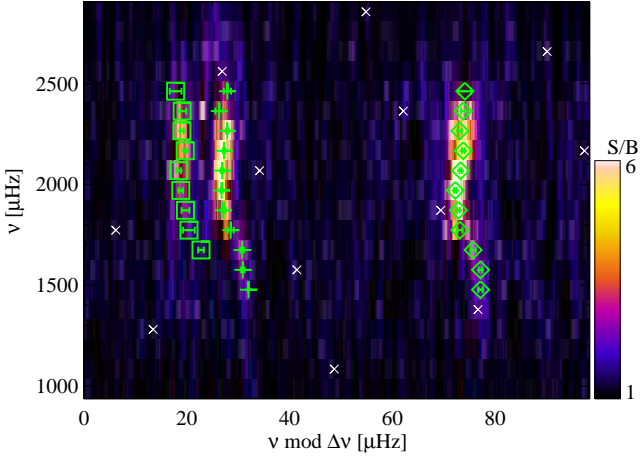}
\caption{\'Echelle diagram of \hd, plotted with a folding frequency $\Delta\nu=98.5\muHz$.
The colour map shows the observed power spectrum normalised by the background (S/B), smoothed over 11 bins. Green symbols with error bars indicate the fitted modes listed in Table~\ref{tab:freq}: plus signs, diamonds, and squares indicate $l=0$, 1, and 2 modes, respectively. White crosses indicate orbital harmonics.}
\label{fig:echelle}
\end{figure}

Around 2000\muHz, the PSD presents a conspicuous comb structure that is typical of solar-like oscillations. A zoom on the PSD in the region is plotted in Fig.~\ref{fig:fit}. The small separation between modes of degrees $l=0$ and 2 is also easily seen, making the mode identification simple. The autocorrelation of the spectrum between 1700 and 2400\muHz\ provides first estimates of the large and small separations: $\Dnu\approx 98.5\muHz$ and $\overline{\delta_{02}}\approx 8\muHz$. These values are recomputed from fitted frequencies in Sect.~\ref{ssec:freq}. We build an \'echelle diagram using \Dnu\ as the folding frequency. This is plotted in Fig.~\ref{fig:echelle}. This \'echelle diagram makes the mode identification even more obvious. On the right-hand side, the clear single ridge corresponds to $l=1$ modes, whereas, on the left-hand side, the two ridges are identified as the $l=2$ and $l=0$ modes. We completely exclude the possibility that the two ridges on the left are $l=1$ modes split by rotation for the following reasons. First, the two ridges show clear asymmetry in their power; second, the rotation needed to generate such high splittings is totally incompatible both with the low-frequency signature (Sect.~\ref{sec:act}) and the spectroscopic observations (Sect.~\ref{ssec:spectro}).
We are able to identify modes presenting a significant signal-to-noise ratio for about ten consecutive orders. Their characteristics are extracted with classical methods, described in the following section.

\subsection{Fitting the spectrum}\label{ssec:fit}
To estimate the mode parameters, we applied usual spectrum fitting techniques, as in previous similar works \citep[e.g.][]{Appourchaux08,Barban09,Deheuvels10}. To summarise the procedure, we assume that the observed PSD is distributed around a mean profile spectrum ${\cal S}(\nu)$ and follows a 2-degree-of-freedom $\chi^2$ statistics \citep[e.g.][]{Duvall86}. ${\cal S}(\nu)$ is modelled as the sum of the background $B(\nu)$, described in Sect.~\ref{sec:bg}, and p-mode profiles ${\cal P}_{l,n}(\nu)$ for each considered degree $l$ and radial order $n$.  ${\cal P}_{l,n}(\nu)$ are multiplets of Lorentzian profiles and read as
\begin{equation}
 {\cal P}_{l,n}(\nu)=\sum_{m=-l}^{l}\frac{a_{l,m}(i) H_{l,n}}{1+[2(\nu-\nu_{l,n}+m\nu_\mathrm{s})/\Gamma_{l,n}]^2},
\end{equation}
where $m$ is the azimuthal order, $H_{l,n}$, $\nu_{l,n}$, and $\Gamma_{l,n}$ are the height, the frequency, and the width of the mode, $\nu_\mathrm{s}$ is the rotational splitting, common to all modes, and $a_{l,m}(i)$, the height ratios of multiplet components, are geometrical terms depending only on the inclination angle $i$ \citep[e.g., see][]{Gizon03}.
Owing to cancellation effects of averaging the stellar flux over the whole disc, only modes with $l \le 2$ are considered in the present analysis.
It is worth noticing that we analyse the full light curve disregarding any possible variations in the p-mode parameters due to any magnetic activity effects as the ones recently uncovered in HD~49933 \citep{Garcia10}.

The different parameters have been independently fitted by ten different groups, using either MLE \citep[e.g.][]{Appourchaux98} or Bayesian priors and performing maximum a posteriori estimation \citep[MAP, e.g.][]{Gaulme09} or MCMC \citep[e.g.][]{Benomar09Bayes,Handberg11}. Generally, global fits of all of the p modes were performed. Nevertheless, one group has used a local approach by fitting sequences of successive $l=2$, 0, and 1 modes, according to the original CoRoT recipes \citep{Appourchaux06}. Between 8 and 15 orders have been included in the fit, depending on the group.

To reduce the dimension of the parameter space, other constraints were used by taking advantage of the smooth variation of heights and widths. As a consequence, only one height and one width, $H_{0,n}$ and $\Gamma_{0,n}$, are fitted for each order.
We linked the height (resp., width) of $l=2$ mode to the height (width) of the nearby $l=0$ mode through a relation 
\begin{equation}
 H_{2,n}=r_2^2 H_{0,n+1}\quad
(\Gamma_{2,n}= \Gamma_{0,n+1}).
\end{equation}
We denote $r_2$ the visibility of the $l=2$ mode relative to $l=0$.
 
Concerning $l=1$ modes, we can link them either to the previous or to the following $l=0$ modes, leading to two fitting situations:
\begin{equation}
\mbox{fit A:}\quad H_{1,n}=r_1^2 H_{0,n+1}\quad
(\Gamma_{1,n}= \Gamma_{0,n+1});
\end{equation}
\begin{equation}
\mbox{fit B:}\quad H_{1,n}=r_1^2 H_{0,n}\quad
(\Gamma_{1,n}= \Gamma_{0,n}).
\end{equation}
where  $r_1$ the visibility of the $l=1$ mode relative to $l=0$.
Even if the variations in heights and widths are limited, they can still be significant over half the large separation. Thus, results can be different for both cases.

The visibility factors $r_l$ depend mainly on the stellar limb-darkening profile \citep[e.g.][]{Gizon03}. It is then possible to fix them to theoretical values, deduced from stellar atmosphere models, or leave them as a free parameters. Six groups have fixed them, whereas four others left them free.

In the next two sections, we present the results of one of the fits used as reference. The reference fitting is based on MLE/MAP, and the fitted frequency range is [1430,2610]\muHz. The background component $B_\mathrm{a}$ is fixed, whereas Bayesian priors for $B_\mathrm{g}$ and $W$ are derived from the background fit (Sect.~\ref{sec:bg}). There are no Bayesian priors for the mode parameters.
The marginal probabilities derived from MCMC derived by other groups are compatible with the error bars presented hereafter.

\subsection{Mode frequencies}\label{ssec:freq}
For the two fitting configurations (A or B), the fitted frequencies are in very good agreement. Table~\ref{tab:freq} provides modes frequencies obtained with fit A. The fitted spectrum is plotted over the observation in Fig.~\ref{fig:fit}. The determined frequencies are also plotted over the \'echelle diagram (Fig.~\ref{fig:echelle}). The frequencies listed in the table were found by at least eight groups out of ten within the error bars. The three modes labelled with an exponent~{\small\itshape a} in the table correspond to modes fitted by only four groups. Nevertheless, the four groups have independently obtained consistent frequencies for these modes, so we report them, but they could be less reliable. An extra $l=1$, around 1455\muHz, seems to be present in the \'echelle diagram and has been fitted by a few groups. Nevertheless, it is very close to one of the orbital harmonics (see Fig.~\ref{fig:echelle}) and has been rejected.

The radial order $n$ provided in Table~\ref{tab:freq} is only relative and could be shifted by $\pm 1$. It is obtained by fitting the relation
\begin{equation}
 \nu_{l,n}=\Dnu(n+l/2+1/4+\alpha)-l(l+1)\overline{\delta_{02}}/6
\end{equation}
derived from the asymptotic development for p modes \citep[e.g.][]{Tassoul80}.

\begin{figure}[!htb]
 \includegraphics[width=\linewidth]{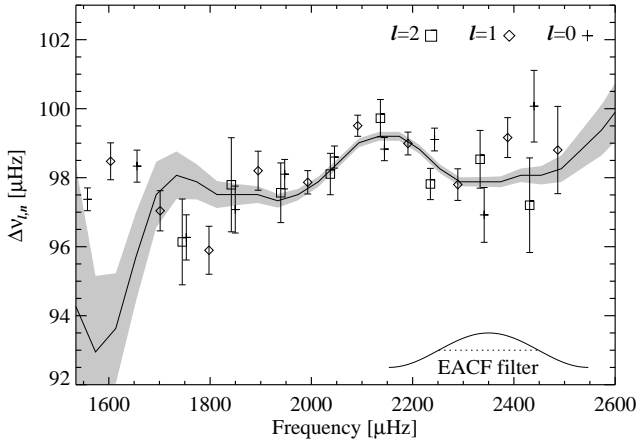}
\caption{Large separations $\Delta\nu_{l,n}=\nu_{l,n}-\nu_{l,n-1}$ for $l=0$, 1, and 2 modes plotted as a function of the frequency $(\nu_{l,n}+\nu_{l,n-1})/2$ for \hd\ (symbols with error bars). A solid line shows the variations in $\Delta\nu$ obtained with the envelope autocorrelation function. The grey area indicates 1-$\sigma$ error bars. The inset shows the cosine filter with a full width at half maximum equal to 2 times the mean large separation used for computing the EACF.}
\label{fig:dnu}
\end{figure}

\begin{table}[!htb]
 \caption{Fitted mode frequencies $\nu_{l,n}$ for \hd.}
\label{tab:freq}\centering
\begin{tabular}{r|*{3}{c@{ \ }r@{$\pm$}c@{}c}} 
\hline\hline
\multicolumn{1}{c|}{$n$}&
$l$&\multicolumn{2}{c}{$\nu_{l,n}$}&&
$l$&\multicolumn{2}{c}{$\nu_{l,n}$}&&
$l$&\multicolumn{2}{c}{$\nu_{l,n}$}& \\
&
&\multicolumn{2}{c}{$\mathrm{[\mu Hz]}$}&&
&\multicolumn{2}{c}{$\mathrm{[\mu Hz]}$}&&
&\multicolumn{2}{c}{$\mathrm{[\mu Hz]}$}& \\
\hline
14 &  0 &  1509.17 & 0.06&\tablefootmark{a}&  1 &  1554.33 & 0.42&\tablefootmark{a}&    & \multicolumn{2}{c}{...}&  \\
15 &  0 &  1606.54 & 0.33&&  1 &  1652.80 & 0.34&&  2 &  1696.88 & 0.53&\tablefootmark{a}  \\
16 &  0 &  1704.88 & 0.33&&  1 &  1749.85 & 0.48&&  2 &  1793.02 & 1.13&  \\
17 &  0 &  1801.15 & 0.56&&  1 &  1845.74 & 0.51&&  2 &  1890.81 & 0.77&  \\
18 &  0 &  1898.22 & 0.38&&  1 &  1943.94 & 0.25&&  2 &  1988.38 & 0.40&  \\
19 &  0 &  1996.32 & 0.20&&  1 &  2041.81 & 0.23&&  2 &  2086.48 & 0.45&  \\
20 &  0 &  2094.92 & 0.25&&  1 &  2141.32 & 0.20&&  2 &  2186.20 & 0.31&  \\
21 &  0 &  2193.75 & 0.23&&  1 &  2240.31 & 0.26&&  2 &  2284.02 & 0.33&  \\
22 &  0 &  2292.86 & 0.24&&  1 &  2338.11 & 0.38&&  2 &  2382.56 & 0.77&  \\
23 &  0 &  2389.78 & 0.76&&  1 &  2437.27 & 0.44&&  2 &  2479.76 & 1.14&  \\
24 &  0 &  2489.85 & 0.71&&  1 &  2536.07 & 1.19&&    & \multicolumn{2}{c}{...}&  \\
\hline
\end{tabular}
\tablefoot{
\tablefoottext{a}{Less reliable frequencies. See text for details.}
}
\end{table}
By using these frequency determinations, we can compute and plot large separations 
\begin{equation}
\Delta\nu_{l,n}=\nu_{l,n}-\nu_{l,n-1}
\end{equation}
 in Fig.~\ref{fig:dnu}. There are clear variations around the mean value $\Dnu=98.3\pm 0.1\muHz$,   recomputed as the average of $\Delta\nu_{l,n}$.
The variation of the large separation can also be measured from the autocorrelation of the time series computed as the power spectrum of the power spectrum windowed with a narrow filter, as proposed by \citet{Roxburgh06} and \citet{Roxburgh09}. \citet{Mosser09ACF} have defined the envelope autocorrelation function (EACF) and explain its use as an automated pipeline. We applied this pipeline to the spectrum of \hd\ to find the frequency variation $\Delta\nu(\nu)$ of the large separation. We used a narrow cosine filter, with a full width at half maximum equal to 2\Dnu. The result is plotted in Fig.~\ref{fig:dnu}. The agreement between $\Delta\nu(\nu)$ and the fitted values $\Delta\nu_{l,n}$ are quite good. We recover the main variations, particularly the large oscillation, visible as a bump around 2100\muHz. Such oscillations in the frequencies and seismic variables can be created by sharp features in the stellar structure \citep[e.g.][]{Vorontsov88,Gough90}. The observed oscillations are probably the signature of the helium's second ionisation zone located below the surface of the star. Such a signature, predicted by models and observed on the Sun, should help us to put constraints on the helium abundance of the star \citep[e.g.][]{Basu04,Piau05}.

\begin{figure}[!htb]
 \includegraphics[width=\linewidth]{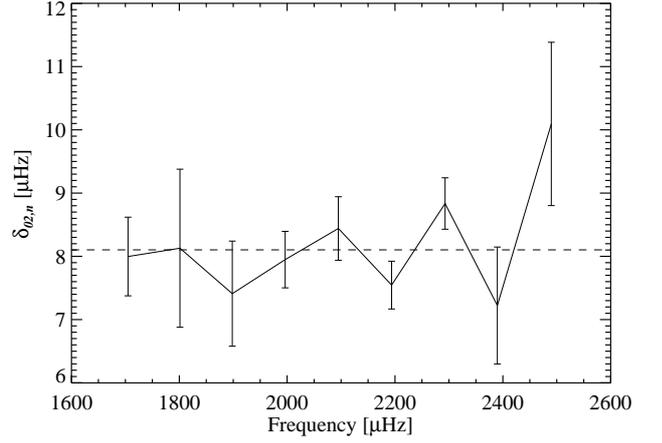}
\caption{Small separations $\delta_{02,n}$ plotted as a function of the frequencies $\nu_{0,n}$ for \hd. The dashed line corresponds to the mean value.}
\label{fig:d02}
\end{figure}

We then computed small separations
\begin{equation}
\delta_{02,n}=\nu_{0,n}-\nu_{2,n-1}. 
\end{equation}
Results are shown in Fig.~\ref{fig:d02}. The error bars take the correlations between the frequency determination of the $l=0$ and 2 modes into account. These correlations are small. It is worth noticing that these small separations do not decrease with frequency as for the Sun, but remain close to their mean value, $\overline{\delta_{02}}=8.1\pm 0.2\muHz$, re-evaluated from fitted frequencies.

\begin{figure}[!htb]
 \includegraphics[width=\linewidth]{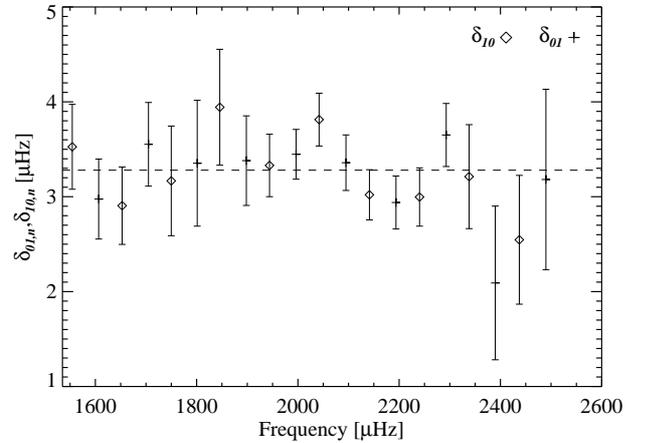}
\caption{Small separations $\delta_{01,n}$ and  $\delta_{10,n}$ plotted as a function of the frequencies $\nu_{0,n}$ and $\nu_{1,n}$ for \hd. The dashed line corresponds to the mean value.}
\label{fig:d01}
\end{figure}

The small separations between $l=0$ and 1 modes, defined by \citet{Roxburgh93} as
\begin{eqnarray}
 \delta_{01,n}&=&\nu_{0,n}-\frac{1}{2}(\nu_{1,n}+\nu_{1,n-1})\quad\mbox{and}\\
 \delta_{10,n}&=&\frac{1}{2}(\nu_{0,n}+\nu_{0,n+1})-\nu_{1,n},
\end{eqnarray}
are also interesting seismic diagnosis tools \citep[e.g.][]{Roxburgh03,Roxburgh05}. Results for \hd\ are shown in Fig.~\ref{fig:d01}. 
As with the $\delta_{02}$ separations they do not display the decrease in frequency 
seen in the solar values, and have an average of $3.28 \pm 0.09\muHz$. These are additional diagnostics of the interior and may even be showing 
an oscillation due to the radius of an outer convective zone and the signature of a convective core
\citep[e.g.][]{Roxburgh09_d01,Silva11}.

The results for the parameters $\nu_\mathrm{s}$ and $i$ obtained by the
different groups are spread but consistent. They appear to be sensitive to
the prior for the background and the frequency range of the fit. A
detailed study dedicated to determining $\nu_\mathrm{s}$ and $i$
will be presented in a separate paper (Gizon et al., in preparation). The
results for the projected splitting $\nu_\mathrm{s}^*=\nu_\mathrm{s} \sin i$,
which is easier to determine than $\nu_\mathrm{s}$ and $i$ separately
\citep{Ballot06}, are also consistent among the different groups. For the
analysis presented in this work we find $\nu_\mathrm{s}^*=0.45 \pm 0.10
\muHz$ for \hd. Moreover, \citet{Ballot08} show that the
frequency estimates are not correlated with estimates for other parameters,
in particular $\nu_\mathrm{s}$ and $i$, so, the frequency estimates are
robust.

\subsection{Lifetimes and amplitudes of modes}

To consider the window effects that affect p modes, we divide all the fitted mode heights by the factor \dc, as mentioned in Sect.~\ref{ssec:window}.
As expected, the values of the fitted heights and widths $H_{0,n}$ and $\Gamma_{0,n}$ change slightly depending on the fitting configuration (A or B) since the values of fitted height and width average the contribution of the modes $(l=0,n)$ and $(l=1,n-1)$ in case A, or of the modes $(l=0,n)$ and $(l=1,n)$ in case B.

\begin{table}[!htb]
 \caption{Fitted widths ($\Gamma$) and amplitudes ($A$) of modes $(l=0,n)$ for \hd, obtained with fits A and B.}
\label{tab:fwhmamp}
\centering
\begin{tabular}{cc r@{}l@{ }c@{ \ }c r@{}l@{ }c@{ \ }c}
\hline\hline
fit& 
$n$ &
\multicolumn{2}{c}{$\Gamma$} & $+$err & $-$err& 
\multicolumn{2}{c}{$A$} & $+$err & $-$err \\
& 
&
\multicolumn{2}{c}{$\mathrm{[\mu Hz]}$} &  & & 
\multicolumn{2}{c}{$\mathrm{[ppm]}$} &  &  \\
\hline
A & 14 &  0.11&\tablefootmark{a}& $+0.22$ & $-0.08$ & 1.34&\tablefootmark{a}& $+0.34$ & $-0.27$ \\
B & 14 &  0.10&\tablefootmark{a}& $+0.03$ & $-0.02$ & 1.24&\tablefootmark{a}& $+0.30$ & $-0.24$ \\
A & 15 &  1.23&& $+1.45$ & $-0.67$ &   1.63&& $+0.29$ & $-0.25$ \\
B & 15 &  1.56&& $+1.25$ & $-0.69$ &   1.81&& $+0.27$ & $-0.23$ \\
A & 16 &  1.45&& $+0.68$ & $-0.46$ &   1.97&& $+0.22$ & $-0.20$ \\
B & 16 &  1.46&& $+0.61$ & $-0.43$ &   2.00&& $+0.23$ & $-0.20$ \\
A & 17 &  2.96&& $+1.31$ & $-0.91$ &   2.38&& $+0.25$ & $-0.23$ \\
B & 17 &  3.87&& $+0.92$ & $-0.74$ &   2.88&& $+0.22$ & $-0.21$ \\
A & 18 &  3.14&& $+0.69$ & $-0.57$ &   3.11&& $+0.22$ & $-0.20$ \\
B & 18 &  2.08&& $+0.55$ & $-0.43$ &   2.99&& $+0.22$ & $-0.21$ \\
A & 19 &  1.53&& $+0.40$ & $-0.32$ &   3.07&& $+0.22$ & $-0.21$ \\
B & 19 &  1.33&& $+0.30$ & $-0.25$ &   3.26&& $+0.23$ & $-0.22$ \\
A & 20 &  1.91&& $+0.38$ & $-0.32$ &   3.72&& $+0.24$ & $-0.22$ \\
B & 20 &  2.03&& $+0.42$ & $-0.35$ &   3.86&& $+0.24$ & $-0.23$ \\
A & 21 &  1.61&& $+0.36$ & $-0.30$ &   3.69&& $+0.24$ & $-0.22$ \\
B & 21 &  1.86&& $+0.38$ & $-0.31$ &   3.68&& $+0.24$ & $-0.22$ \\
A & 22 &  1.92&& $+0.38$ & $-0.31$ &   3.70&& $+0.23$ & $-0.22$ \\
B & 22 &  2.29&& $+0.42$ & $-0.36$ &   3.74&& $+0.23$ & $-0.22$ \\
A & 23 &  4.02&& $+0.76$ & $-0.64$ &   3.58&& $+0.22$ & $-0.20$ \\
B & 23 &  5.08&& $+1.15$ & $-0.94$ &   3.51&& $+0.22$ & $-0.21$ \\
A & 24 &  4.39&& $+1.21$ & $-0.95$ &   2.84&& $+0.21$ & $-0.20$ \\
B & 24 &  5.88&& $+2.56$ & $-1.78$ &   2.30&& $+0.24$ & $-0.22$ \\
\hline
\end{tabular}
\tablefoot{
\tablefoottext{a}{Less reliable determinations. See text for details.}
}

\end{table}
\begin{figure}[!htb]
 \includegraphics[width=\linewidth]{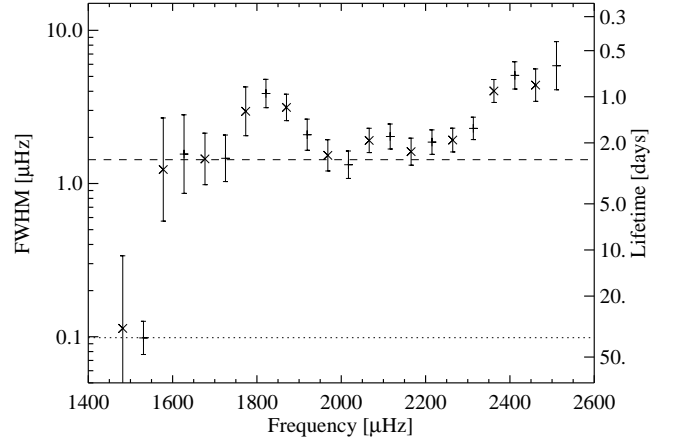}
\caption{Full width at half maximum of modes as a function of frequency for \hd. Crosses correspond to values obtained by fitting the same widths for modes $(l=1,n-1)$ and $(l=0,n)$, i.e. fit A, whereas plus signs correspond to values obtained by fitting the same widths for modes $(l=1,n)$ and $(l=0,n)$, i.e. fit B. Widths are plotted as a function of the mean frequency of the two modes. The dotted line indicates the spectral resolution. The dashed line shows predictions by \citet{Chaplin09}. The y-axis on the right-hand side shows the corresponding lifetimes.}
\label{fig:fwhm}
\end{figure}

The mode width is a direct measurement of the mode lifetime $\tau_{0,n}=(\pi\Gamma_{0,n})^{-1}$. The widths of fitted modes are listed in Table~\ref{tab:fwhmamp} and plotted in Fig.~\ref{fig:fwhm}. The two lowest values are quite low and close to the spectral resolution. Moreover, the values of fitted widths obtained for these modes by the different groups are rather widespread, compared to the other widths, which are consistent. As a consequence, these two widths are not considered as reliable and must be rejected.
If we exclude these two points, the mode widths correspond to lifetimes ranging from $\sim$0.5~days at the highest frequency to $\sim$3~days at the lowest. The variation of widths is not monotonic and shows an S shape. The widths progressively increase until a local maximum of $\sim$4\muHz\ ($\tau\sim 1$ days) around 1850\muHz\ corresponding to the mode $(l=1,n=17)$ that actually appears to be significantly wider than its neighbours. Then, the widths decrease until a plateau of $\sim$2\muHz\ ($\tau\sim 2$ days) covering $\sim$4 orders, and finally increase again.

\hd\ mode lifetimes are
shorter than the solar ones, but significantly longer than lifetimes previously observed in F stars, such as HD~49933 \citep[e.g.][]{Appourchaux08}. Mode lifetimes clearly decrease as the effective temperature increases. By using theoretical models and former observations, \citet{Chaplin09} find that the mean lifetimes of the most excited modes scale as $\teff^{-4}$, which leads to $\tau\approx2.6$~days for $\teff=6100$~K. This value is indicated in Fig.~\ref{fig:fwhm}. We notice a qualitative agreement with the observations, but the lifetimes in the plateau are all shorter than the value deduced from this scaling law.
Recently, \citet{Baudin10} have used a sample of more accurate seismic observations -- including these observations of \hd\ -- to show that the dependence of lifetimes on the effective temperature is even stronger and that it scales with $\teff^{-14}$.

\begin{figure}[!htb]
 \includegraphics[width=\linewidth]{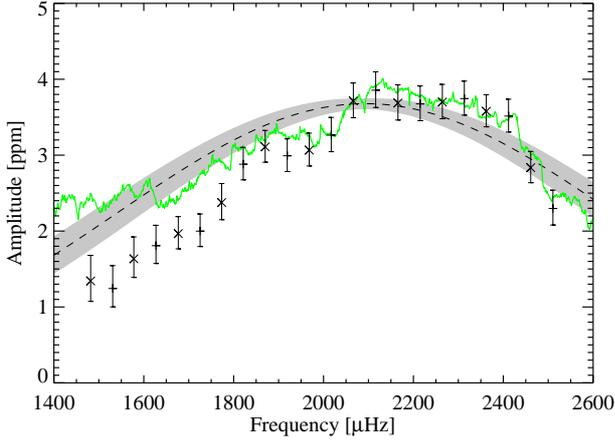}
\caption{Amplitudes of radial modes as a function of frequency for \hd.  Crosses correspond to values obtained by fitting the same intrinsic amplitudes for modes $(l=1,n-1)$ and $(l=0,n)$, i.e. fit A, whereas plus signs correspond to values obtained by fitting same intrinsic amplitudes  for modes $(l=1,n)$ and $(l=0,n)$, i.e. fit B. Amplitudes are plotted as a function of the mean frequency of the two modes. The dashed line shows the amplitudes of radial modes deduced from the fit as a Gaussian profile of the p-mode power excess (see Sect.~\ref{sec:bg}). Grey area indicates the associated 1-$\sigma$ error bars. The green curve shows the amplitudes of radial modes deduced from the spectrum smoothed by a \Dnu-wide boxcar, after subtracting the background.}
\label{fig:amp}
\end{figure}

From $H_{0,n}$ and $\Gamma_{0,n}$, it is possible to recover the rms amplitude $A_{0,n}$ of a mode. Amplitudes are always better determined than heights and widths themselves. This is true regardless of the values of $i$ and $\nu_\mathrm{s}$ \citep{Ballot08}.
The amplitudes follow the relation
\begin{equation}
 A_{0,n}= \sqrt{\frac{\pi}{2}\Gamma_{0,n} H_{0,n}.}
\end{equation}
The mode amplitudes are listed in Table~\ref{tab:fwhmamp} and plotted in Fig.~\ref{fig:amp}. The errors take the correlations between the determinations of $H_{0,n}$ and $\Gamma_{0,n}$ into account. The amplitudes increase almost regularly until reaching a maximum (3.86 ppm) around the frequency 2100\muHz. Then, the mode amplitudes remain close to this maximum before sharply dropping above 2450\muHz.

It is also possible to derive the amplitude of radial modes from the smoothed power spectrum after subtracting the background \citep[see][]{Kjeldsen08}, or, similarly, by using the fitted Gaussian profile $B_\mathrm{p}(\nu)$ of the p-mode power excess derived in Sect.~\ref{sec:bg}.
By using \citet{Michel09}, we recover the amplitude of radial modes through a relation
\begin{equation}
 A=\frac{R_{l=0}}{R_\mathrm{osc}}\sqrt{B_\mathrm{p}\Dnu},
\end{equation}
where $R_{l=0}$ and $R_\mathrm{osc}$ are the CoRoT response functions for $l=0$ modes and for the sum of all modes of degree $l=0$ to 4 \citep[see definitions in][notice that $R_{l=0}$ is identical to the granulation response $R_g$ defined in that article]{Michel09}. We used the approximated formulations of the responses they derived, and find $R_{l=0}=3.89$ and $R_\mathrm{osc}=6.86$, by considering $\teff\approx6100$~K for \hd. 

When the spectrum is barely resolved, this procedure is a common way to extract the mode amplitudes. In the present situation, we are thus able to compare these estimates to the individually fitted mode amplitudes. In Fig.~\ref{fig:amp}, we have also plotted the radial mode amplitudes deduced from the smoothed spectrum and from the profile $B_\mathrm{p}$.
Although the amplitudes of radial modes do not follow a perfect Gaussian profile, the Gaussian profile is a reasonable fit to the smoothed spectrum.
The Gaussian profile reaches the maximum value $\displaystyle A^\mathrm{(max)}=\frac{R_{l=0}}{R_\mathrm{osc}}\sqrt{H_\mathrm{p}^\mathrm{(max)}\Dnu}\approx 3.7$~ppm. This is consistent with the maximum fitted amplitude ($A_{l=0}^{(\mathrm{max})}=3.86\pm0.24$~ppm). 
We get very good agreement at high frequency, but there is a clear departure at low frequency between the amplitudes obtained from the smoothed spectrum and from the fitted modes. 
The fitted background is probably underestimated around $\sim$1200--1800\muHz. This could come from a component missing in our model, like a contribution of faculae or from an excess of power due to leakage effects.
Thus, when the granulation is fitted in Sect.~\ref{sec:bg}, the extra power increases the apparent contribution from p modes; however, when we fit p modes, in a restricted range  (Sect.~\ref{ssec:fit}), the extra power excess is included in the background.

The amplitudes are measured in the CoRoT spectral band and must be converted into bolometric amplitudes. According to \citet{Michel09}, the bolometric correction factor is $c_\mathrm{bol}=4/R_{l=0}$ and is  $c_\mathrm{bol}=1.03$ for \hd. The maximum bolometric amplitude of radial mode is then
\begin{equation}
 A_{l=0,\mathrm{bol}}^{(\mathrm{max})}=3.96\pm0.24\unit{ppm}.
\end{equation}

By combining the adiabatic relation proposed by \citet{Kjeldsen95} to relate mode amplitudes in intensity to mode amplitudes in velocity and the scaling law proposed by \citet{Samadi07}, we get the relation
\begin{equation}
 A^{(\mathrm{max})}=
A^{(\mathrm{max})}_{\sun}
\left(
\frac{L/L_{\sun}}{M/M_{\sun}}
\right)^{0.7}
\sqrt{ \frac{\teff}{T_{\mathrm{eff},\sun}} }.\label{eq:scalamp}
\end{equation}
Moreover, if we assume that the frequency of maximum p-mode amplitude $\nu^{(\mathrm{max})}$ scales with the acoustic cut-off frequency \citep[e.g., see][]{Bedding03}, we obtain the relation \citep{Deheuvels10}:
\begin{equation}
 A^{(\mathrm{max})}=
A^{(\mathrm{max})}_{\sun}
\left(\frac{\teff}{T_{\mathrm{eff},\sun}}\right)^{1.95}
\left(\frac{\nu^{(\mathrm{max})}}{\nu^{(\mathrm{max})}_{\sun}}\right)^{-0.7}.
\end{equation}
By using
$A_{l=0,\mathrm{bol},\sun}^{(\mathrm{max})}=2.53\pm0.11\unit{ppm}$ \citep{Michel09},
$T_{\mathrm{eff},\sun}=5777$~K, 
and $\nu^{(\mathrm{max})}_{\sun}=3050\muHz$ for the Sun 
and by considering for \hd\ our estimations $\teff=6100\pm60$~K, 
and $\nu^{(\mathrm{max})}=2090\pm20\muHz$, 
we obtain $A_{l=0,\mathrm{bol}}^{(\mathrm{max})}=3.7\pm0.2\unit{ppm}$.
This predicted value is slightly lower than the value measured for \hd, but still consistent within the error bars. In contrast to previous observations of F stars \citep[e.g.][]{Michel08} which have smaller amplitudes than expected, the observations for this G0V star are close to predictions.
 
The mode visibility $r_1$ and $r_2$ are free parameters in the reference fitting. The values obtained are given in Table~\ref{tab:vis} for fits A and B. Both results are consistent. 
These values are compatible with theoretical values computed with the CoRoT limb-darkening laws by \citet{Sing10}. As a further test, we considered 
the sum of visibilities of modes over an interval \Dnu\  that is equal to $1+r_1^2+r_2^2$. The fitted values are listed in Table~\ref{tab:vis}. According to \citet{Michel09}, this is comparable to the quantity $(R_\mathrm{osc}/R_{l=0})^2=3.10$. The second includes modes up to $l=4$, whereas the first includes modes only up to $l=2$. Nevertheless, the agreement between fits A and B and the theoretical expectation is met within the error bars.

\begin{table}[!htb]
 \caption{Fitted amplitude mode ratios $r_l=A_l/A_0$ compared to theoretical values.}\label{tab:vis}
\centering
\begin{tabular}{cccc}
\hline
\hline
& fit A & fit B & theo.\\
\hline
 $r_1$ & $1.22\pm0.06$ & $1.21\pm 0.06$ & 1.22\\
 $r_2$ & $0.76\pm0.05$ & $0.74\pm 0.05$ & 0.72\\
$1+r_1^2+r_2^2$ & $3.07\pm0.19$ & $3.02\pm 0.19$ & 3.01\\
\hline
\end{tabular}

\end{table}

\section{Conclusion}\label{sec:conclusion}
Using photometric observations from the CoRoT space telescope spanning 117 days and a duty cycle of 90\%, we measured a rotation period $P_\mathrm{rot}=12.3\pm0.15$ days for the G0 main sequence star \hd, thanks to a modulation in the light curve induced by photospheric activity. We have clearly detected solar-like oscillations in \hd, and characterised 31 p modes in the range 1500--2550\muHz\ with degrees $l=0$, 1, and 2. We observed lifetimes for these modes ranging between 0.5 and 3 days. \hd\ mode lifetimes are shorter than the solar ones, but significantly longer than the lifetimes previously observed in F stars, confirming that mode lifetimes decrease as the effective temperature increases.
Moreover, we observed for \hd\ that the variation in lifetimes with frequency is not monotonic and shows a clear S shape. 

The fitted maximum bolometric amplitude for radial modes is $3.96\pm0.24$~ppm, which is marginally higher than the theoretical models of \citet{Samadi07} but still compatible within the error bars. In the past, several analyses have shown smaller amplitudes than predicted; however, they were F stars, whereas \hd\ is a G0 star, hence more like the Sun. Nevertheless, this star is over metallic. \citet{Samadi10a} have 
shown that the surface metallicity has a strong impact on the 
efficiency of the mode driven by turbulent convection: the lower the 
surface metal abundance, the weaker the driving.
These authors have precisely quantified this effect for the 
CoRoT target HD 49933, which is a rather metal-poor star compared to the 
Sun since for this star $\feh=-0.37$.  \citet{Samadi10b} found 
that ignoring the surface metal abundance of this target results in a 
significant underestimation of the observed mode amplitudes.
The theoretical scaling law by \citet{Samadi07} (see Eq.~\ref{eq:scalamp}) was 
obtained on the basis of a series of 3D hydrodynamical models with a 
solar metal abundance. 
Given the result of \citet{Samadi10b,Samadi10a}, we would expect for \hd\ higher theoretical mode amplitudes than predicted by the 
theoretical scaling law of \citet{Samadi07}. However, the amount by 
which the theoretical amplitudes are expected to increase remains to be 
precisely quantified and compared to the uncertainties associated with the 
present seismic data.

For \hd, we found that the mean large and small separations are $\Dnu = 98.3 \pm 0.1\muHz$ and $\overline{\delta_{02}} = 8.1\pm 0.2\muHz$ and that $\delta_{02}$ does not significantly decrease with frequency. These quantities are typical of a $\sim$1.2-solar-mass star that is still on the main sequence \citep[see preparatory models by][]{Soriano07}. Moreover, \hd\ does not show any mixed modes, as is the case for the more evolved G0 star HD~49385 \citep{Deheuvels10}. 

The variation of $\Delta\nu$ with frequency shows an oscillation that we interpret as a possible signature of the second helium-ionisation region. Thanks to accurate eigenfrequency measurements  (the error is about 0.2\muHz\ for the modes with the highest amplitudes) and to fine measurements of fundamental parameters obtained with the Narval spectrograph, \hd\ is a very promising object for stellar modelling. These observations should especially help in determining whether \hd\ has a convective core or not, which would put useful constraints on the mixing processes in such stars.
These seismic inferences will also help improve our knowledge of the planet hosted by \hd.

\begin{acknowledgements}
JB acknowledges the support of the Agence National de la Recherche through the SIROCO project.
IWR, GAV, WJC, and YE acknowledge support from the UK Science and Technology Facilities Council (STFC).
This work has been partially supported by the CNES/GOLF grant at the Service d'Astrophysique (CEA/Saclay) and the grant PNAyA2007-62650 from the Spanish National Research Plan. NCAR is supported by the National Science Foundation. 
This work benefited from the support of the International Space Science Institute (ISSI), by funding the AsteroFLAG international team. 
It was also partly supported by the European Helio- and Asteroseismology Network (HELAS), a major international collaboration funded by the European Commission's Sixth Framework Programme, and by the French PNPS programme.
This research made use of the SIMBAD database, operated at the CDS, Strasbourg, France.
\end{acknowledgements}

\bibliographystyle{aa}
\bibliography{biblio}

\end{document}